\documentclass[twocolumn,superscriptaddress]{revtex4}
\usepackage{graphicx}
\usepackage{epstopdf}
\usepackage{amsmath}
\usepackage{booktabs}
\usepackage{color}
\usepackage{multirow}
\begin{document}

\title{Importance of many-body correlations in glass transition:\\an example from polydisperse hard spheres} 

\author{Mathieu Leocmach$^a$}
\email{mathieu.leocmach@polytechnique.org}
\altaffiliation[Present address: ]{Laboratoire de Physique, CNRS UMR 5672, Ecole Normale Supérieure de Lyon, 46 allée d'Italie 69364 Lyon cedex 07, France.}
\author{John Russo$^a$}
\email{russoj@iis.u-tokyo.ac.jp}
\thanks{\\$^a$M. Leocmach and J. Russo contributed equally to this work.}
\author{Hajime Tanaka}
\email{tanaka@iis.u-tokyo.ac.jp}
\affiliation{ {Institute of Industrial Science, University of Tokyo, 4-6-1 Komaba, Meguro-ku, Tokyo 153-8505, Japan} }
\date{Received October 12, 2012}

\begin{abstract}
Most of the liquid-state theories, including glass-transition theories, are constructed on the basis of two-body density correlations. 
However, we have recently shown that many-body correlations, in particular bond orientational correlations, play a key role in
both the glass transition and the crystallization transition.
Here we show, with numerical simulations of supercooled polydisperse hard spheres systems, that the lengthscale associated with any two-point spatial 
correlation function does not increase toward the glass transition. A growing lengthscale is instead revealed
by considering many-body correlation functions, such as correlators of orientational order,
which follows the lengthscale of the dynamic heterogeneities.
Despite the growing of crystal-like bond orientational order, we reveal that the stability against crystallization with increasing polydispersity
is due to an increasing population of icosahedral arrangements of particles. 
Our results suggest that, for this type of systems, many-body correlations are a manifestation of the link between the vitrification and the crystallization phenomena.
Whether a system is vitrified or crystallized can be controlled by the degree of frustration against crystallization, polydispersity 
in this case.  
\end{abstract}

\maketitle

\section{Introduction}

Amorphous materials have been of prime importance in our technology for millennia, from antique glass works to fashionable phones made of metallic glass. One of the new frontiers of the amorphous technology is in the design of amorphous drugs~\cite{Petit2006,Grzybowska2012}, better absorbed by our metabolism with less side effects, that would be stable at room temperature. The main obstacle is a lack of our basic understanding of the physics of the glass transition, without any operative consensus despite half a century of intensive research~\citep{cavagna2009supercooled,BerthierR}.

When cooled below its freezing temperature while avoiding crystallization, a liquid becomes supercooled. Upon further cooling, the dynamics slows down by many orders of magnitude leading to a material that is mechanically a solid without long range positional order, thus called amorphous. It is now known that the dynamics in a supercooled liquid is heterogeneous, with a length scale that grows when approaching the glass transition~\citep{yamamoto1998,Donati1999a} (see~\citep{BerthierR} for a review). The lengthscale defined by the dynamical heterogeneity is not static (one-time spatial correlation) but dynamic (two-time spatial correlation).

The existence of a static (structural) length that would grow and accompany the dynamic heterogeneities is still not clear in the general case. However, in a class of system that includes polydisperse hard spheres, we have shown~\cite{tanaka} that some medium range bond orientational order reminiscent of the crystal exists in the supercooled liquid and grows toward the glass transition in the same way as the dynamical heterogeneity. The presence in glassy materials of structures locally reminiscent of crystals has been confirmed recently in amorphous silicon~\cite{Treacy2012} and in a metallic glass~\cite{Hwang2012}. While bond orientational order is a member of the class of many-body
correlations between neighboring particles, it is yet unclear if a similar lengthscale can be extracted
from two-body correlation functions.
This question is particularly important considering that the Mode-Coupling Theory (MCT) of the glass transition takes as input only two-body quantities, and
similarly modern spin-glass-type theories of the structural glass transition~\cite{lubchenko2007,Biroli2008,Parisi2010} are not taking explicitly into account many-body correlations.

At polydispersities over $6 \sim 7$ \%, the system needs to fractionate to crystallize~\citep{Fasolo2003}. What is the bond order of the reference crystal is then unclear and may challenge our scenario. This is a situation reminiscent of binary hard sphere systems of size ratio close to one~\cite{Hopkins2011b,Hopkins2012} where growing bond order have not been reported~\cite{Charbonneau}. However, it is known that even in binary systems locally favored structures play a role in the slowing down of the dynamics in some cases~\cite{Coslovich2011,Malins2012}. Interestingly, recent studies by Mosayebi et al. \cite{mosayebi2010,mosayebi2012} demonstrate that there are static growing lengths 
in binary Lennard-Jones and soft sphere mixtures. They estimated the static length by looking at the spatial correlation of the degree of non-affine deformation 
of inherent structures under shear deformation. 

In the present study we will use polydisperse hard sphere systems, where we know how to extract meaningful many-body correlations, and look for a two-body quantity that would show the same behavior as the bond order. We will show that the two-body part of the free energy (which for hard potentials corresponds to the two-body part of the structural entropy) is unable to capture medium range bond ordered regions or to yield correlation lengths meaningful from the point of view of the glass transition. We thus confirm the medium range crystalline order scenario and test its robustness against increasing polydispersity.
Since bond orientational order is directly linked to the underlying crystalline structures, we will then address the important question
of what is the mechanism responsible for the avoidance of crystallization. We will show that in the metastable fluid phase crystalline packings
are in competition with icosahedral packings, and that polydispersity acts in favor of the latter ones.

The paper is organized as follows. In Sec.~\ref{sec:methods} we present the details of the simulations and the order parameters considered in this work. In Sec.~\ref{sec:results} the results are organized into
a study of the order parameters distribution (Sec.~\ref{sec:order_parameters}) and their static lengthscales (Sec.~\ref{sec:lengths}), and a method to determine the competition between crystalline structures
and icosahedral packings (Sec.~\ref{sec:icosahedra}). 
In Sec.~\ref{sec:discussion} we discuss our results. In Sec.~\ref{sec:summary} we summarize our work. 

\section{Methods}\label{sec:methods}

\subsection{Simulation method} \label{sec:methods_sim}
We run isothermal-isobaric NpT Monte Carlo simulations of $N=4000$ polydisperse hard spheres.
The diameters ($\sigma$) follow a Gaussian distribution $P(\sigma)=\exp{\left[-(\sigma-\sigma_0)^2/2\,s^2\right]}/\sqrt{2\pi} s$,
with polydispersity index $\Delta=s/\sigma_0$. In the following we fix the unit of length as $\sigma_0=1$ and the unit
of energy so that the Boltzmann constant is unity, $k_B=1$.

\subsection{Estimation of two-point quantities: pair entropy and pair free energy} \label{sec:methods_two}
Our aim is to compare the behavior of both two-point quantities and many-body quantities with increasing pressure.
Due to the hard-sphere interaction, entropy is the only contribution to the system free energy.
All two-pair correlation quantities are thus derived from the two-body excess entropy~\cite{Nettleton1958,Mountain1971},
defined as
\begin{equation}
s_2=-\frac{\rho k_B}{2}\int dr\left[g(r)\log(g(r))-g(r)+1\right]. 
\end{equation}

In principle, $s_2$ can be calculated separately for each particle $i$ in the system. In practice, this requires time averages to compute the pair correlation function of each particle, $g_i(r)$, where the particle distribution around a particle $i$ is averaged over short-time scales ($\beta$ processes). In Refs. \cite{tanaka,watanabe_walls,KawasakiJPCM} we averaged on times comparable or longer than the $\alpha$ relaxation, leading to a quantity that was trivially a reflection of the dynamical heterogeneity. 

Here we instead construct an approximate but instantaneous $s_2(i)$ using the pair correlation function $g(r)$
\begin{equation}
s_2(i) = -\frac{\rho k_B}{2}\sum_j \left[g(r_{ij})\log(g(r_{ij}))-g(r_{ij})+1\right].
\end{equation}
This quantity is in very good agreement with the one obtained by calculating the radial distribution function for each
particle, $g_i(r)$, averaged over times comparable to the $\beta$ relaxation time.

More rigorously, one can compute directly the free-energy of each configuration by
measuring the free volume of the particles, defined as the volume ($v(i)$) in which each sphere can freely
move while holding all the other spheres fixed. It has been shown~\cite{Aste2004} that this free
volume is simply related to the pair free-energy ($f_2$) by the following relation
\begin{equation}
f_2=\sum_i f_2(i)=-k_BT\sum_i \log(v(i)/\lambda),
\end{equation}
where $\lambda$ is the thermal de Broglie wavelength. To compute the free volume $v(i)$ we follow previous
studies~\cite{Aste2004}: first the Voronoi-diagram for each configuration is computed, and the polyhedron
surrounding each particle is determined. To account for polydispersity we employ the radical Voronoi tessellation.
The free volume of particle $i$ is then
computed by shifting normally all the faces of the corresponding polyhedron by $\sigma(i)/2$
toward particle $i$, and computing the new volume. In this way
the volume $v(i)$ represents the volume in which the excluded volume of particle $i$ can move without leaving its Voronoi cell.
This procedure is conducted independently
for each particle and for each configuration.

\subsection{Estimation of many-point quantities: bond orientational order parameter analysis}\label{sec:methods_many}
To study many-body correlations we use the local bond-order analysis introduced by
Steinhardt~\cite{steinhardt}, first applied to study crystal nucleation by
Frenkel and co-workers~\cite{auer}. 
The $\ell$-fold symmetry of a neighborhood around each particle $i$ is characterized by a $(2\ell+1)$ dimensional complex vector ($\mathbf{q}_l$) as $q_{\ell m}(i)=\frac{1}{N_b(i)}\sum_{j=1}^{N_b(i)} Y_{\ell m}(\mathbf{\hat{r}_{ij}})$, where
$\ell$ is a free integer parameter, and $m$ is an integer
that runs from $m=-\ell$ to $m=\ell$. The functions $Y_{\ell m}$ are the spherical harmonics
and $\mathbf{\hat{r}_{ij}}$ is the vector from particle $i$ to particle $j$.
The sum goes over all neighboring particles $N_b(i)$ of particle $i$. Usually 
$N_b(i)$ is defined by all particles within a cutoff distance, but in an inhomogeneous system
the cutoff distance would have to change according to the local density. Instead we
sort neighbors according to their distance from particle $i$, and
fix $N_b(i)=12$ which is the number of nearest neighbors in icosahedra and close packed crystals (like \textsc{hcp} and \textsc{fcc})
which are known to be the only relevant crystalline structures for hard spheres.

In the analysis, one uses the rotational invariants defined as:
\begin{align}
	q_\ell =& \sqrt{\frac{4\pi}{2l+1} \sum_{m=-\ell}^{\ell} |q_{\ell m}|^2 }, \label{eq:ql}\\
	w_\ell =& \sum_{m_1+m_2+m_3=0} 
			\left( \begin{array}{ccc}
				\ell & \ell & \ell \\
				m_1 & m_2 & m_3 
			\end{array} \right)
			q_{\ell m_1} q_{\ell m_2} q_{\ell m_3}, \label{eq:wl}
\end{align}
where the term in brackets in Eq.~(\ref{eq:wl}) is the Wigner 3-j symbol.
In particular both crystalline and icosahedral neighborhood have high $q_6$ (strong 6-fold symmetry), with the highest values for the latter. To detect specifically icosahedral order one prefers $w_6$, whose minimum value is obtained only by a perfect icosahedron.

The identification of crystalline particles follows the usual procedure~\cite{auer}. A particle is identified as crystal if
its orientational order is coherent (in symmetry and in orientation) with that of its neighbors.
The scalar product $(\mathbf{q}_6(i)/|\mathbf{q}_6(i)|)\cdot(\mathbf{q}_6(j)/|\mathbf{q}_6(j)|)$ quantifies this similarity. If it exceeds $0.7$ between
two neighbors, they are deemed \emph{connected}. We then identify a particle as crystalline if it is connected with at least $7$ neighbors~\cite{auer}. In a more continuous way, summing the contribution of all the bonds of a given particle, we define the ``crystallinity''~\cite{russo_hs}
\begin{equation}\label{eqn:crystallinity}
 \text{C}(i)=\sum_{j=0}^{N_b(i)}\frac{\mathbf{q}_6(i)\cdot\mathbf{q}_6(j)}{|\mathbf{q}_6(i)|\,|\mathbf{q}_6(j)|}.
\end{equation}

Alternatively, one can coarse-grain $\mathbf{q}_\ell$ over the neighbors~\cite{lechner}
\begin{equation}
	Q_{\ell m}(i) = \frac{1}{N(i)+1}\left( q_{\ell m}(i) +  \sum_{j=0}^{N(i)} q_{\ell m}(j)\right), 
	\label{eq:Qlm}
\end{equation}
to suppress the signal from locally incoherent orientations (icosahedral order)~\cite{mathieu_icosahedra} and use the resulting invariant $Q_6$ as an indication of crystallinity, more precisely, the degree of crystal-like rotational symmetry. Alternatively, we note that the shortcomings of
non-coarse-grained order parameters in the identification of crystallinity can be addressed by Minkowski tensors~\cite{kapfer2012jammed}.

Here we briefly consider the physical meaning of crystal-like bond orientational order parameters.  
The crystallization transition is characterized by
the symmetry breaking of both orientational and translational order.
We note that both $\text{C}$ and $Q_6$ are good measures of bond orientational order, whereas the density or other two-body
quantities are measures of translational order.  It was shown~\cite{russo_hs} that in hard spheres crystallization is
driven by fluctuations in bond orientational order and not by density fluctuations. Crystals continuously form, grow and melt
in regions of high bond orientational order, which then effectively act as precursors for the crystallization transition.
So $\text{C}$ and $Q_6$, while not
being direct indicators for the presence of crystals, rather measure the tendency to promote crystallization. 
In Sec.~\ref{sec:lengths} we are going to show
that indeed the lengthscale associated with bond orientational order fluctuations increases with supercooling. Then in Sec.~\ref{sec:icosahedra} we are going to study
the mechanism by which the crystallization transition is avoided.

\subsection{Estimation of the correlation lengths of various quantities} \label{sec:methods_length}
Finally we explain how to evaluate the correlation length of various order parameters. The calculation can be carried out both
in real space and in Fourier space. While formally containing the same information, the Fourier analysis has some practical advantages over
the real space analysis.
In real space, the correlation function of any order parameter is a oscillating and rapidly decaying function of $r$. The
correlation length is obtained by fitting the \emph{envelope} of the correlation function with a Ornstein-Zernike expression.
This expresion is only asymptotic, so the two first peaks at short $r$ should be omitted, and it is also rapidly decaying, so that
the statistical noise strongly affects the quality of the fit at long $r$. The problem is less severe for order parameters
having a tensorial nature, and correlation lengths of crystal-like bond orientational order have been easily measured in real space in previous studies~\cite{tanaka,kawasaki,mathieu_icosahedra,russo_gcm,russo_hs}. For example, the tensorial order parameter $Q_{6m}$ effectively
correlates 7 scalar fields, allowing a sevenfold reduction of the noise. But the real space analysis requires much longer
time averages for two-body correlation functions, both $f_2$ and $s_2$, which have a purely scalar nature.
This problem can be overcome by calculating the correlation functions in Fourier space. These function are not oscillatory at small $q$ where we can expect a Ornstein-Zernike form, and thus much easier to fit unambiguously.
So, to keep both two-body and many-body
correlation functions consistent, we calculate all correlations in Fourier space.
We explain in the following a straightforward procedure to extract correlation length from Fourier space analysis.

At a given time step, for any scalar order parameter field $x$ increasing with order, we compute a structure factor keeping only the 10\% most ordered particles (more on this choice below). This condition defines a threshold $x^*$. The ensemble average of the thresholds $\langle x^*\rangle$ are indicated on Fig.~\ref{fig:distrib}. Formally we define a function $\omega(i) = \Theta [x(i) - x^*]$, where $\Theta(x)$ is the Heaviside’s step function, and a four-point structure factor
\begin{equation}
	S_x(q) = N^{-1}(\left\langle \Omega(\mathbf{q}) \Omega(-\mathbf{q}) \right\rangle - | \left\langle \Omega(\mathbf{q}) \right\rangle^2 |),   
	\label{eq:StrutureFactor}
\end{equation}
where $\Omega(\mathbf{q})$ is the Fourier transform of $\omega(i)$: 
\begin{equation}
	\Omega(\mathbf{q}) = \sum_i \omega(i)\exp(-\imath \mathbf{q}\cdot\mathbf{r}_i).  
\end{equation}
This structure factor is then ensemble averaged (still noted $S_x(q)$ for concision) over $\approx 10^4$ configurations. The case of order parameters decreasing toward ordering (i.e. $s_2$, $w_6$) is trivially obtained by changing the sign.

\section{Results}\label{sec:results}

\begin{figure}
 \centering
 \includegraphics{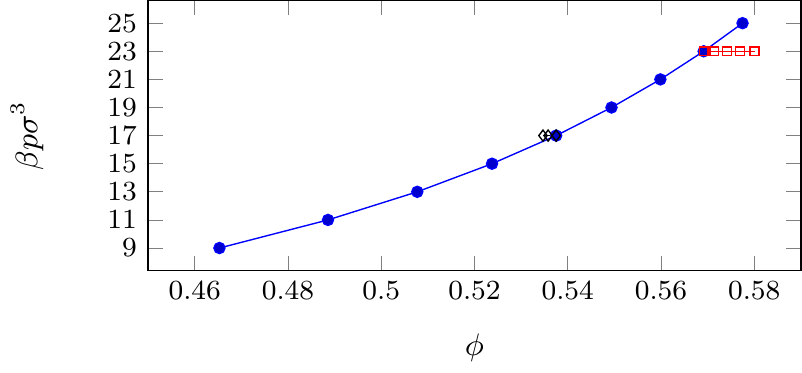}
 \caption{Simulated state points. The blue filled circles (with blue curve) represent the equation of state for polydispersity $\Delta=7\%$. The red open squares (with red line) instead are simulation points at the same pressure ($\beta p\sigma^3=23$) but at different polydispersities $\Delta$: from low to high volume fraction they correspond to $\Delta=7$\%, 9\%, 11\%, 13\%, and 15\% respectively. Similarly the black open diamonds (with black line) are at $\beta p\sigma^3=17$ and $\Delta=0$\%, 4\%, and 7\%.}
 \label{fig:eos}
\end{figure}

Figure~\ref{fig:eos} shows the equation of state for the simulated state points. In particular
we consider three data-sets. The first one (blue filled circles in the figure) corresponds to
simulations at a constant polydispersity of $\Delta=7\%$. For each state point we run $8$ independent
simulation runs and extract configurations for the calculation of correlation lengths.
The second data set (red open squares in the figure) are instead isobaric simulation (at $\beta p\sigma^3=23$)
with increasing polydispersity, $\Delta=7\%,9\%,11\%,13\%,15\%$. The third data set (black open diamonds in the figure) are also isobaric simulations at $\beta p\sigma^3=17$ with increasing polydispersity, $\Delta=0\%,4\%,7\%$. The two last data sets are used to study
the mechanism by which crystallization is suppressed upon an increase of polydispersity, unveiling the
role played by icosahedral arrangement of particles.

\subsection{Order parameter distribution and mobility}\label{sec:order_parameters}

\begin{figure}
	\centering
	\includegraphics{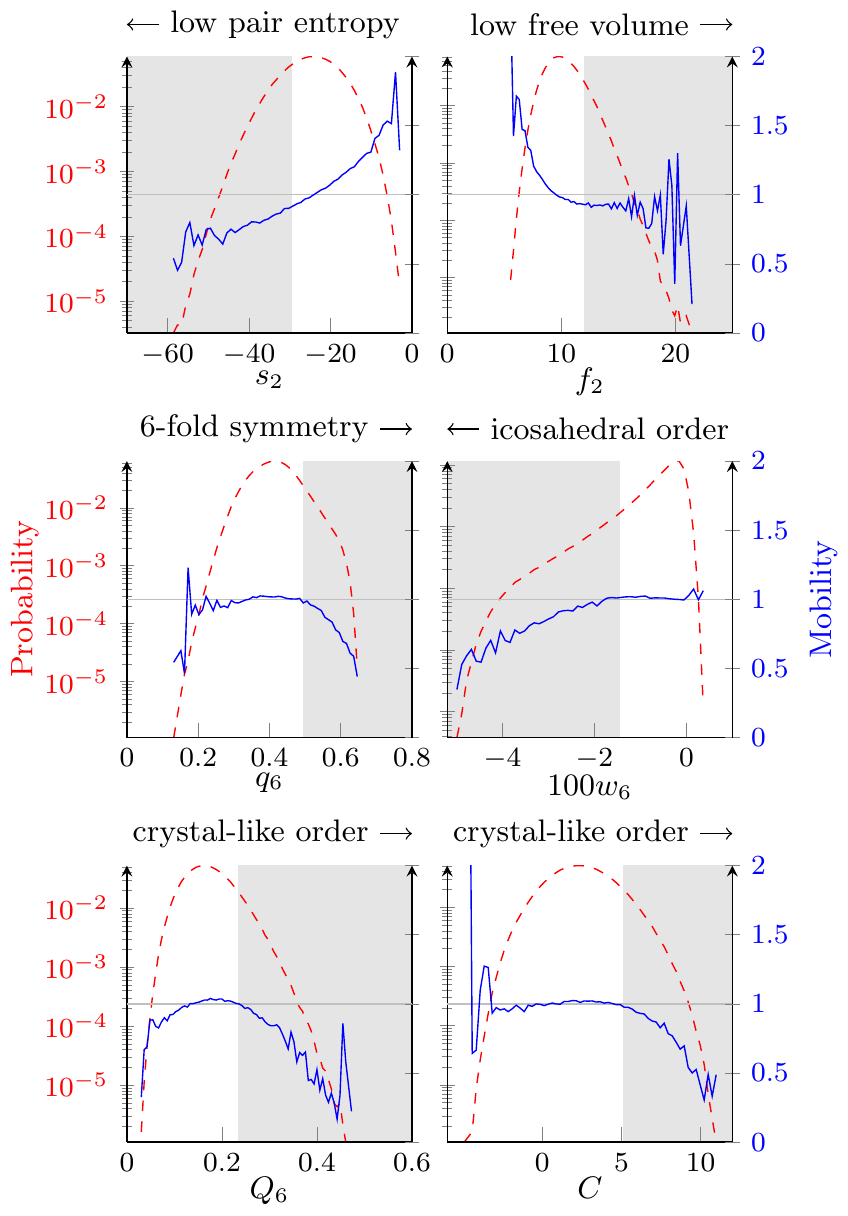}
	\caption{Probability distributions (red dashed line) and mobility (blue continuous line) function of various order parameters at $\beta p\sigma^3=25$ and for a time difference corresponding to the $\alpha$-relaxation. Top row: two-body excess entropy $s_2$ (left) and pair free energy $f_2$ (right). Central row: local six-fold orientational orders $q_6$ (left) and $w_6$ (right). Bottom row: coarse grained $Q_6$ (left) and crystallinity $C$ (right). Mobility is in unit of mean-square displacement. The shaded area shows the contribution from $10\%$ of the particles having highest (for $f_2$, $q_6$, $Q_6$ and $C$) or lowest (for $s_2$, $w_6$) value of the order parameter.}
	\label{fig:distrib}
\end{figure}

We study systematically two-body ($s_2$, $f_2$) and many-body ($q_6$, $w_6$, $Q_6$, $C$) scalar order parameter fields for our highest pressure ($\beta p\sigma^3=25$). For each of these parameters one can define if a particle is ``ordered'' or not. Very negative values of $s_2$ indicate low two-body configurational entropy and thus \emph{a priori} some kind of ordering or stability. Similarly, high values of $f_2$ indicate high two-body free energy (low free volume). High values of $Q_6$ or $C$ indicate locally crystalline environment (locally similar to \textsc{hcp} or \textsc{fcc} crystal), whereas very negative values of $w_6$ correspond to icosahedral packings. High values of $q_6$ can indicate ambiguously either crystalline or icosahedral environments.

In Fig.~\ref{fig:distrib} we show the ``ordered'' side of each parameter as a shaded area, and the probability distribution of this parameter as a red dashed line. Note that for any parameter, the maximum probability do not correspond to the ordered side. However the probability distribution decays more slowly on the ordered side, indicating the presence of rare but very ordered particles.

\begin{figure}
 \centering
 \includegraphics{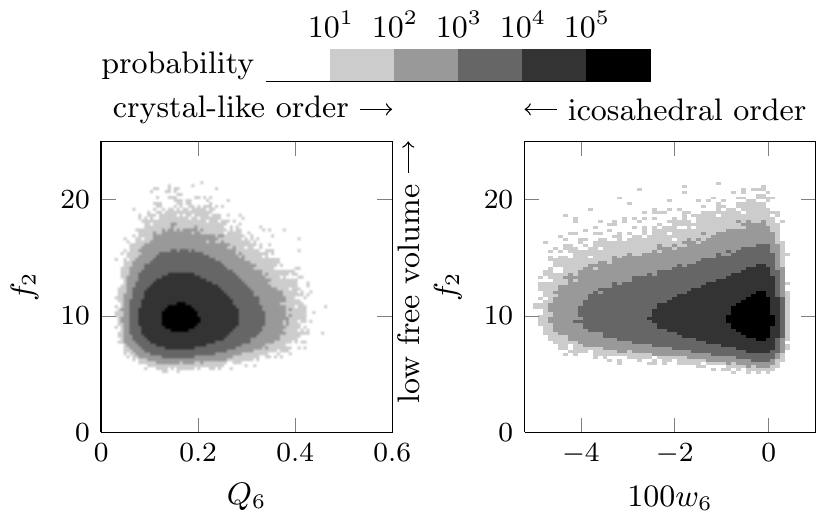}
 \caption{Correlation between two-body and many-body parameters. Probability distribution functions in the $f_2$-$Q_6$ map (left) and in the $f_2$-$w_6$ map (right) for a metastable fluid with polydispersity $\Delta=7\%$ and pressure $\beta p\sigma^3=23$. There is no linear correlation between the parameters, meaning
 that high values of $Q_6$ (the crystalline regions) or low values of $w_6$ (the icosahedral regions) are not extremum values of $f_2$.}
 \label{fig:f2decoupling}
\end{figure}

By plotting the probability distribution function of the metastable fluid in the $(Q_6, f_2)$ and $(w_6, f_2)$ planes, Fig.~\ref{fig:f2decoupling} shows the absence of linear correlations between $f_2$ and both $Q_6$ and $w_6$. Since $Q_6$ identifies regions of high crystal-like bond orientational order and $w_6$ locates icosahedral arrangements of particles, it is clear that high $f_2$ regions are not associated with any of these structures. We checked in the same way that $s_2$ is not correlated with many-body parameters.

The absence of strong correlations between two-body quantities and both crystalline and icosahedral packings is also evident from
the microscopic dynamics.
To study the correlation between any scalar order parameter $x$ and the displacement of the particles, we define the mobility 
\begin{equation}
	\Delta r^2(x=x_0, t) \equiv \left\langle \frac{
		\sum\limits_i{
			\left\|\mathbf{r}_i(t)-\mathbf{r}_i(0)\right\|^2 \delta(x(i)-x_0)
			}
	}{
		\sum\limits_i{\delta(x(i)-x_0)}
	}\right\rangle,
	\label{eq:mobility}
\end{equation}
shown in Fig.~\ref{fig:distrib} for a time difference corresponding to the $\alpha$-relaxation time. Note that we use $\delta$ functions of a constant finite width and thus the number of particles involved in the average of Eq.~(\ref{eq:mobility}) varies like the probability distribution, explaining the noise in the low probability regions.

We found that for each parameter, its mobility decreases with increasing order. The mobilities of bond-order quantities are flat in the disordered regions and decrease when approaching the perfect structure (i.e. icosahedron for $w_6$, crystal structures for $Q_6$ or $C$). By contrast, the mobility of two-body order parameters tends to increase strongly in the disordered region and decrease less in the ordered region (it is almost flat at high $f_2$). We conclude that many-body quantities describe better the slowing down accompanying good local ordering, while two-body quantities are not clearly correlated to such slower structures.
Note that in Fig.~\ref{fig:distrib} both icosahedral packings (low $w_6$ particles) and bond-ordered crystalline regions (high $Q_6$ and $C$) are associated with slow dynamics. As was shown by some of us~\cite{mathieu_icosahedra}, the structures primarily responsible for the slowing down of the dynamics are the crystal-like particles, while icosahedral particles act to prevent the crystal nucleation process~\cite{russo_hs}.
This will be shown in detail in Sec.~\ref{sec:icosahedra}.

\subsection{Length-scales}\label{sec:lengths}
\begin{figure}
	\centering
	\includegraphics{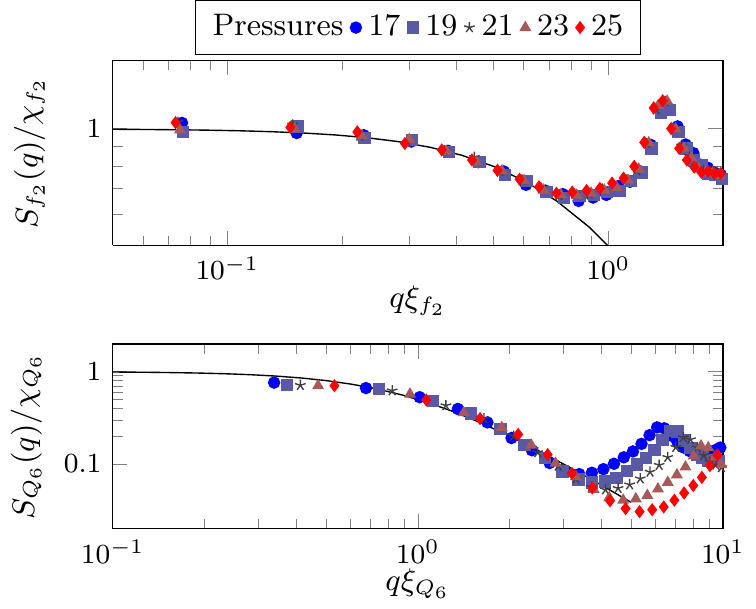}
	\caption{Structure factors collapse on the Ornstein-Zernike form for (top) $f_2$ and (bottom) $Q_6$. Note that the $f_2$ structure factors are almost similar for all pressures considered. 
	For $Q_6$ instead, the growth of the correlation length is evident already from the systematic change in the relative position of the nearest-neighbors peaks. }
	\label{fig:structurefactor}
\end{figure}

\begin{figure}
	\centering
	\includegraphics{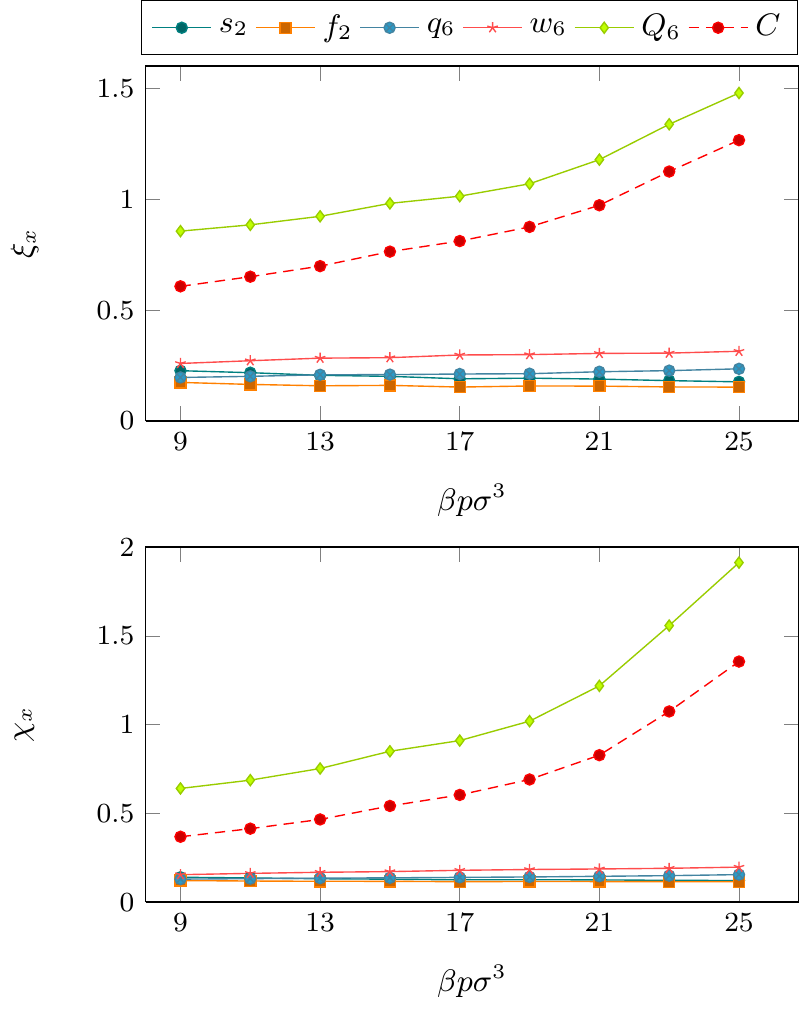}
	\caption{Correlation length ($\xi_x$) and susceptibilities ($\chi_x$) extracted for two-body and many-body scalar order parameters, 
plotted as a function of the pressure. Only many-body correlation lengths are increasing (only slightly for $\xi_{w_6}$), while the lengthscale associated with the two-body quantities is almost constant.}
	\label{fig:Fourierlengths}
\end{figure}

\begin{figure}
	\centering
	\includegraphics{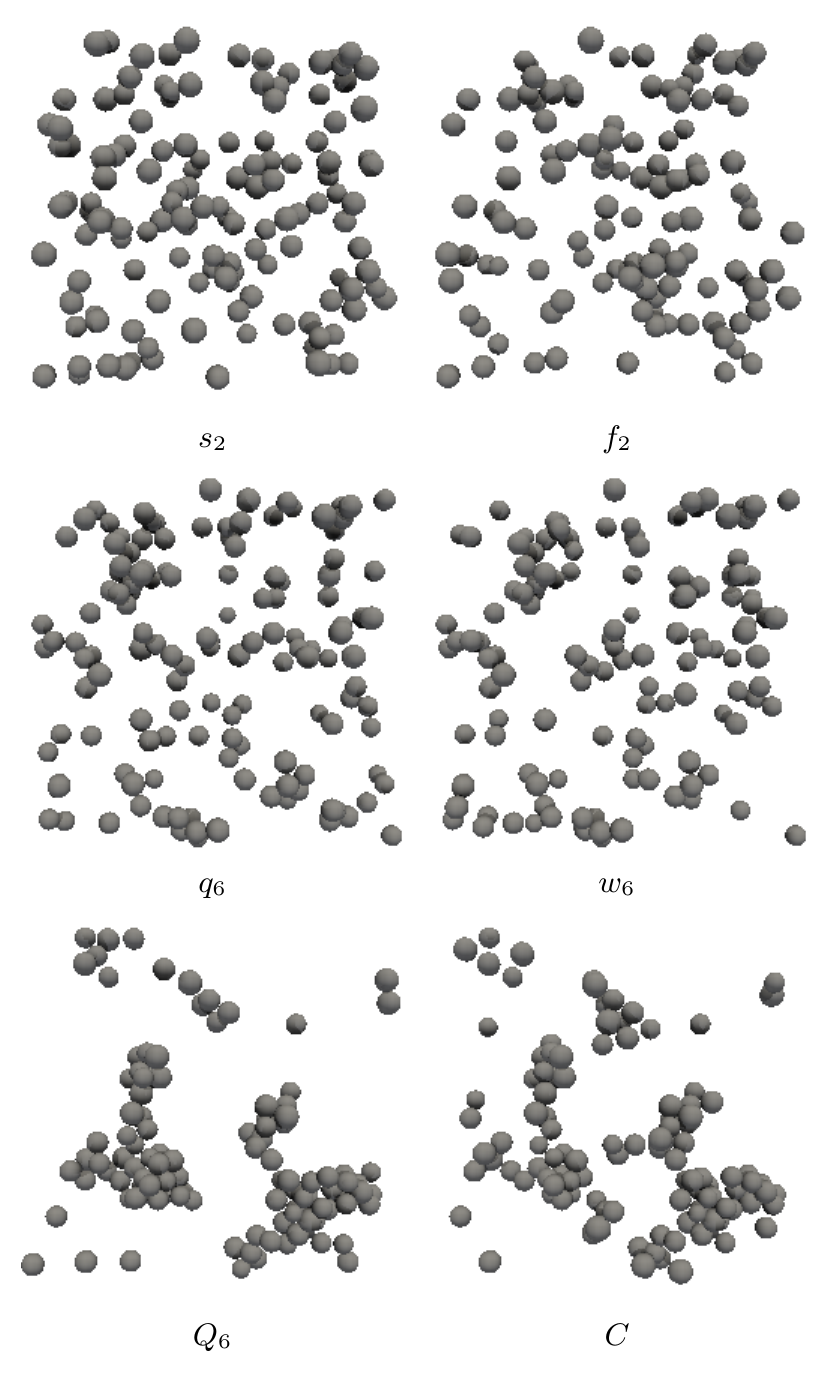}
	\caption{Visualization of the 10\% most ordered particles defined by the various order parameters. All pictures correspond to a thin slice ($5\sigma$) of the same configuration at $\beta p\sigma^3=23$ and $\Delta=7\%$. Only $Q_6$ and $C$ show meaningful spatial fluctuations. We can also see anti-correlation between ($Q_6$,$C$) and ($q_6$,$w_6$). }
	\label{fig:3D}
\end{figure}

As explained in Sec.~\ref{sec:methods_length}, we estimate the correlation lengths of various quantities $x$ in Fourier space 
not only for the two-body $s_2$ and $f_2$, but also for scalars derived from the multi-body bond orientational order, i.e. $q_6$, $w_6$, $Q_6$ and $C$, which allows us to have overall coherency of the length-scale analysis of all the parameters. 

Figure~\ref{fig:structurefactor} shows the increase of $S_x(q)$ toward small wavenumbers ($x=f_2$ upper panel; $x=Q_6$ lower panel), which is well described by the asymptotic Ornstein-Zernike function in Fourier space:
\begin{equation}
	S_x(q\rightarrow 0) \approx \frac{\chi_x}{1+\xi_x^2 q^2},
	\label{eq:OZ_Fourier}
\end{equation}
where $\xi_x$ is the correlation length and $\chi_x$ the susceptibility of fluctuations of quantity $x$. In general, an independent determination of $\chi_x$ is crucial for the fit~\cite{Flenner2011}. However here we deal with correlation lengths much smaller than the simulation box and both $\xi_x$ and $\chi_x$ can be reliably estimated from a two parameter fit of $S_x(q)$. We note that the absence of the finite-size effects was confirmed for this situation \cite{tanaka}.  

The dependence on the pressure of the resulting correlation lengths $\xi_x$ is shown in Fig.~\ref{fig:Fourierlengths}. Most of the order parameters produce constant lengthscales, including not only the two-body quantities but also $q_6$, $w_6$ that are sensible to icosahedral order. The only growing lengths are extracted from measures of local crystal-like order, i.e. $Q_6$ and $C$. We confirm that the same results are obtained from real-space correlation functions (but in
real space is possible to detect the growth also in $q_6$, see~\cite{nota_q6}).

The absence of correlations for two-body quantities and the presence of a growing lengthscale for $Q_6$ and $C$ are evident also by direct inspection of the particle configurations.
Figure~\ref{fig:3D} plots the $10\%$ most ordered particles for the different order parameters at $\beta p\sigma^3=23$ and $\Delta=7\%$. The first row of Fig.~\ref{fig:3D} shows the absence of any
appreciable correlation for both two-body quantities $f_2$ and $s_2$. The middle row shows that also the signal from icosahedral clusters
(both $q_6$ and $w_6$ have icosahedra as their extremum) display no appreciable correlation length, i.e. they form randomly and homogeneously throughout the system. 
Only $Q_6$ and $C$ (last row in Fig.~\ref{fig:3D}) show clustering of the ordered particles on medium range.

The length scales obtained from $Q_6$ or $C$ in Fourier space and from the tensorial $\mathbf{q}_6$ or $\mathbf{Q}_6$ in real space (not shown) are similar and increase monotonically with pressure. Note that the scalar $q_6$ (dominated by the icosahedral order) yields a very different correlation length. This is consistent with the spatial coherency (in orientation) of crystal-like order that is missing in icosahedral order~\cite{tanaka,mathieu_icosahedra}. Coherently, real-space correlation function (not shown) of $f_2$ (respectively $s_2$) are perfectly identical at all pressures. 

The choice of the threshold $x^*$ is a balance between taking in too many particles or too few. If too few (below $5\%$) $S_x$ is too noisy. If too many, the threshold does not discriminate between ordered and disordered particles and $S_x$ tends to the trivial density $S(q)$. We found that between $5\%$ and $40\%$ the absolute value of the length depends marginally on $x^*$ but its pressure dependence does not. We chose to use $10\%$ across this paper because this value allows the easiest direct visualization on a single frame (Fig.~\ref{fig:3D}). We checked the pressure independence of the two-body parameter's length with thresholds up to $90\%$.

The study of correlation lengths has shown that by increasing pressure, the range of
crystal-like bond orientational order increases, driving the slowing down of the system as shown in Refs.~\cite{tanaka,mathieu_icosahedra}. 
Bond orientational order corresponds to
orientationally ordered regions which spontaneously form in the metastable phase.
$f_2$ is instead decoupled from the relevant structures involved in the transition, as
was shown in Fig.~\ref{fig:f2decoupling}. We now address the question of how the system avoids crystallization
despite the growing lenghtscale of bond orientational order.

\subsection{Competition between icosahedral arrangements and crystal-like arrangements}\label{sec:icosahedra}

\begin{figure}
 \centering
 \includegraphics{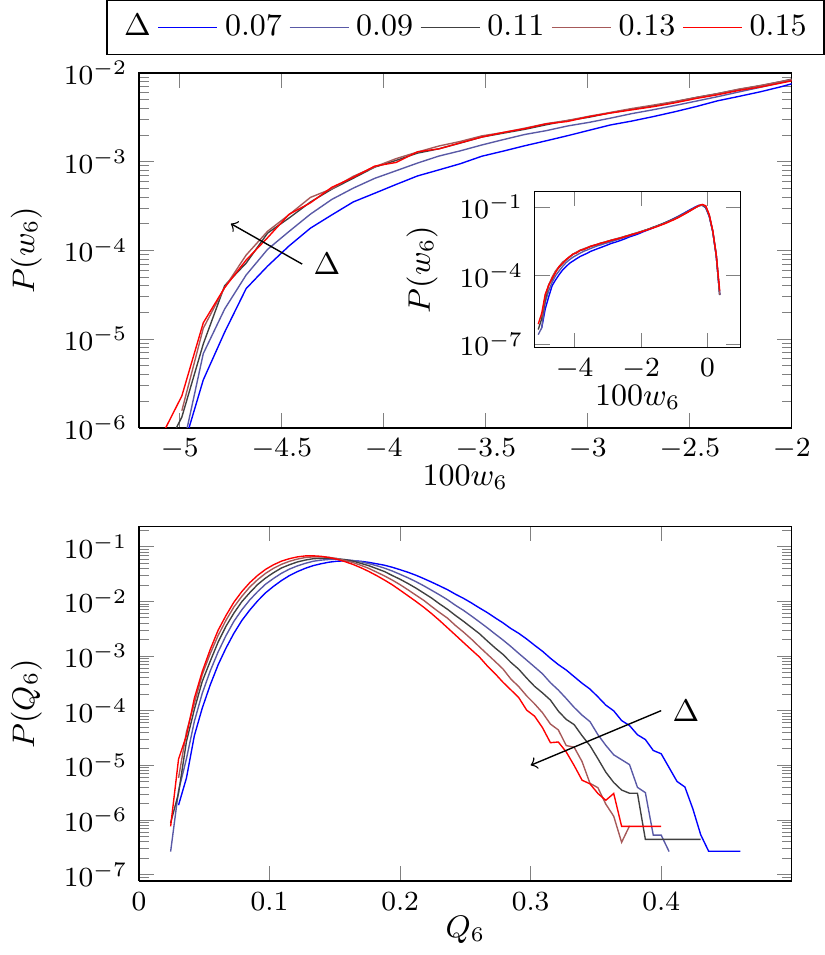}
 \caption{Effect of polydispersity on local structures at constant pressure ($\beta p\sigma^3=23$): (top) detail of the distribution of $w_6$ (full distribution in inset) showing a small increase in the icosahedra population saturating around $10\%$; (bottom) distribution of $Q_6$ showing a marked decrease in the crystallinity with polydispersity.}
 \label{fig:polydispersity}
\end{figure}

We will now focus on the state point at $\beta p\sigma^3=23$ at different polydispersities to
study the mechanism by which polydispersity disfavors the crystallization transition.

In Fig.~\ref{fig:polydispersity} we show the probability distributions for the order
parameters $Q_6$ and $w_6$ at different polydispersities. It is immediately
evident that, while bond orientational order is rapidly suppressed with increasing
polydispersity (as shown in the suppressed signal at high $Q_6$), particles in icosahedral
environments are not disfavored by polydispersity. On the contrary the fraction of
icosahedral particles increases with polydispersity, and saturates at around $\Delta=10\%$.
This is in agreement with recent evidence of increased icosahedral ordering with size disparity
in metallic glasses~\cite{Shimono2012}. Intuitively, we can conclude that icosahedral order is more tolerant to size
asymmetry (with the small particle usually sitting at the center of the icosahedral cage) than crystalline order is.

\begin{figure}
 \centering
 \includegraphics{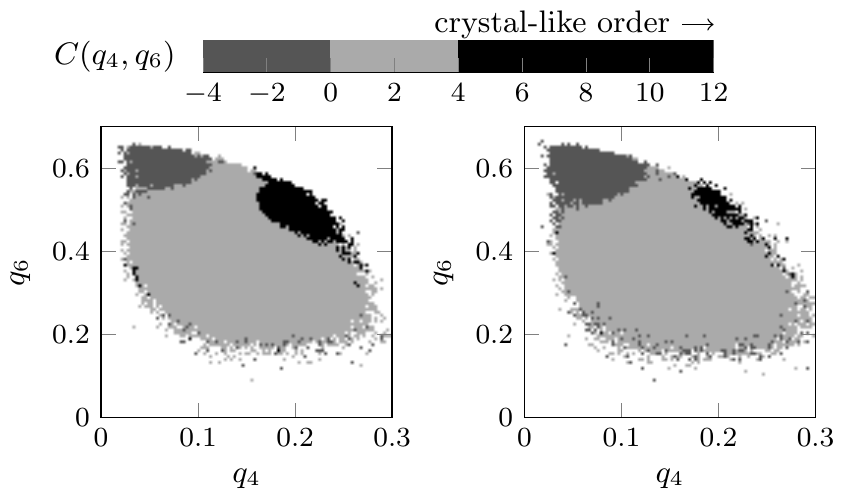}
 \caption{Average crystallinity order parameter projected on the $q_4$-$q_6$ plane for the metastable fluid at $\beta p\sigma^3=23$ at $\Delta=7\%$ (left) and $\Delta=15\%$ (right). Icosahedra appear in the top-left corner of each plot (in dark gray) and perfect \textsc{fcc} crystals would be in the top-right corner with $C=12$ (in black).}
 \label{fig:Cmaps}
\end{figure}

Figure~\ref{fig:Cmaps} shows that the metastable fluid distribution on the $q_4$-$q_6$ plane, which is a convenient
representation as perfect icosahedral packings sit on the top-left corner of the distribution, while perfect crystals on
the top-right corner. The figure clearly shows that, by increasing polydispersity, the biggest change in the metastable fluid
distribution is the suppression of crystal-like regions (high values of
$q_4$, $q_6$ and $C$, the black region), while icosahedral environments are slightly enhanced (high $q_6$, low $q_4$ and negative $C$, the dark gray region).

\begin{figure}
 \centering
 \includegraphics{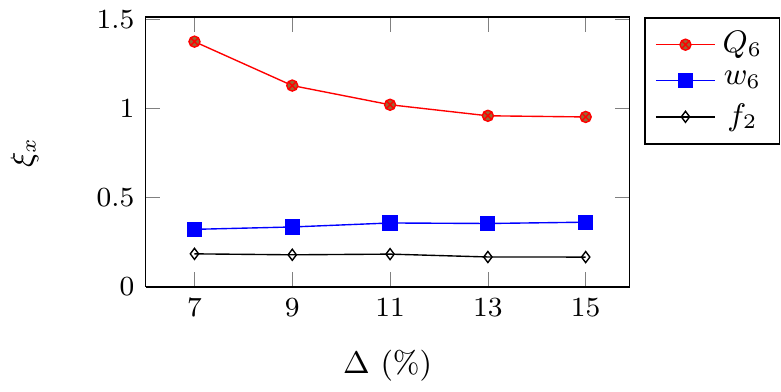}
	\caption{Polydispersity dependence of the correlation lengths at $\beta p\sigma^3=23$. The dominant crystalline length decreases, the icosahedral length increases, however they plateau well before crossing. The two-body length shows no indication of becoming dominant with increasing $\Delta$, even decreasing slightly.}
	\label{fig:lengthpoly}
\end{figure}

The different effects of polydispersity on icosahedral and crystalline ordering are reflected in the different correlation lengths.
Figure~\ref{fig:lengthpoly} shows the correlation length extracted from $Q_6$, $w_6$ and $f_2$ as a function of polydispersity. While the correlation length for $Q_6$ (associated with crystal-like regions) decreases with increasing polydispersity, the one extracted from $w_6$ (associated with icosahedral regions) increases.
However the two lengths are far from crossing and seem to saturate around $\Delta=13-15\%$. The correlation length of $f_2$ is constant or even slightly decreasing with $\Delta$, never taking over the many-body lengths. It is thus clear that, in the present range of polydispersity, crystal-like bond order fluctuations is still the dominant contribution in the static (and dynamic~\cite{mathieu_icosahedra}) properties of the system.

\begin{figure}
 \centering
 \includegraphics{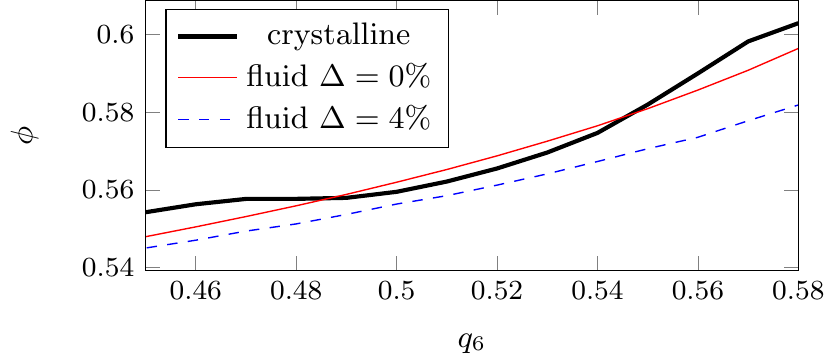}
 \caption{Average $\phi$ as a function of $q_6$ for particles identified as ``fluid'' and ``crystalline'' according to the criteria outlined in Sec.~\ref{sec:methods}C. The continuous red curve represents fluid particles in a system at $\beta p\sigma^3=17$ and $\Delta=0\%$, while dashed blue curve represents fluid particles at the same pressure but at $\Delta=4\%$. The thick black curve represents instead crystalline particles (this curve is less sensitive to polydispersity and it is reported once).
 Note that, while in the monodisperse case there is a $\phi$ interval in which the crystalline particles have higher orientational order, in the polydisperse
 case crystalline particles have lower orientational order at all $\phi$. While the monodisperse case is easily crystallized in direct simulations, the polydisperse
 system always remains metastable.}
 \label{fig:stability_map}
\end{figure}

In Ref.~\cite{russo_hs} we have shown that the competition between crystalline packings and
icosahedral packings can be studied via two-dimensional maps of translational vs orientational order.
Orientational order is captured by $q_6$, which is small for disordered arrangement of particles and
increases for both crystal-like and icosahedral particles. Translational order is instead measured with
the local packing fraction, $\phi$, obtained by measuring the volume of the Voronoi diagram associated with
each particle. The calculation is straightforward: for each
configuration, crystalline particles are identified with the method described in Sec.~\ref{sec:methods_many} and the other particles are termed ``fluid''. For each subset of particles, the average value of the volume fraction $\phi$ is calculated as a function of the order parameter $q_6$, and plotted in Fig.~\ref{fig:stability_map}.
For each value of $q_6$ the map captures the average volume fraction $\phi$ of both crystalline and non-crystalline
environments.

In Fig.~\ref{fig:stability_map}
we compare the curves at $\beta p\sigma^3=17$ and at different polydispersities (see also the black diamonds of Fig.~\ref{fig:eos}): 
at $0\%$ (monodisperse system, red continuous curve) and at $4\%$ (blue dashed curve). The curve for the
crystalline particles is similar at the two considered polydispersities and is reported once
as the thick solid line. First we consider the monodisperse case. As shown in the figure,
at low volume fraction, a particle in the fluid phase has higher orientational order than in the crystalline phase.
But at $\phi\cong 55.8\%$ a crossover
occurs and the crystal phase gains microscopic stability: a particle in a crystalline environment will
have higher orientational order than a particle in the fluid phase at the same volume fraction.
This crossover marks the appearance in the fluid phase of the metastable crystals which continuously
appear, grow and shrink, until eventually a crystal droplet reaches the critical size and starts the
crystallization process.
At a higher volume fraction ($\phi\cong 58\%$) another crossover occurs, with
crystalline particles having less orientational order than particles in the ``fluid'' branch.
These particles in the fluid phase at high $q_6$ are easily identified as particles in icosahedral packings (they have
a low value of $w_6$).
It is thus clear that icosahedral packings are competing with crystalline packings, eventually dominating
at high $\phi$. This scenario is confirmed by looking at the polydisperse case (blue dashed line in
Fig.~\ref{fig:stability_map}).
At polydispersity $\Delta=4\%$ the crystalline branch always lays above the fluid one, meaning that
crystalline environments, at any fixed $\phi$, are not able to attain higher bond orientational order than other configurations.
Microscopically the difference between the monodisperse and the polydisperse case is due to an increased population of icosahedral particles, which dominate the fluid branch for $q_6\gtrsim 0.5$.
We have shown that, while for the monodisperse case there is a range of volume fractions where crystalline particles
attain higher orientational order than icosahedral arrangements, for the polydisperse case icosahedral
particles always attain higher orientational order.
This is immediately reflected in direct simulations,
as at this pressure it is not possible to crystallize simulations at $\Delta=4\%$, while the
monodisperse simulations are easily crystallized~\cite{zaccarelli,pusey2009hard}. Since the diffusional dynamics of the
system, which controls the kinetic factor of crystallization \cite{TanakaK}, 
is approximately the same at different values of $\Delta$~\cite{zaccarelli}, this difference in the crystallization behavior has to come from
some structural difference introduced by polydispersity, i.e. an increased population of icosahedral particles (see Figs. \ref{fig:polydispersity} and \ref{fig:Cmaps}).
We also confirm that at polydispersity $\Delta=7\%$ icosahedral particles are favored over crystalline ones for
all the pressures studied, in line with the observation of Fasolo and Sollich~\cite{fasolo2004fractionation} that
one-phase crystallization is suppressed at high polydispersity.

We have thus provided direct evidence that icosahedral particles are responsible for the suppression
of crystallization in polydisperse hard spheres.

\section{Discussion} \label{sec:discussion}

In the present study we have compared the evolution of static length scales in polydisperse hard spheres
for both two-body and many-body correlation functions. While two-body correlation functions do not
show any sign of an increasing length scale with pressure, we have confirmed that in the glass transition of polydisperse hard spheres the relevant static length is the correlation length of the crystal-like bond order.
We have also determined the relevant microscopic structures that are associated with the increasing
lengthscale: they correspond to crystal-like environments of particles and are characterized by slow
dynamics.
We thus confirm that in the glass transition of polydisperse hard spheres the relevant static length is the correlation length of the crystal-like bond orientational order.

The other relevant structure with slow dynamics are icosahedral packings of particles, but
their lengthscale does not grow appreciably with increasing pressure. Icosahedral assemblies of particles
are spatially uncorrelated. While not having a direct role in the slowing down of the dynamics, we
have shown that icosahedral packings are responsible for the avoidance of the crystallization transition
with increasing polydispersity.
The increase in polydispersity reduces the degree of crystal-like bond orientational order whereas enhances the icosahedral order (see Figs. \ref{fig:polydispersity} and \ref{fig:Cmaps}). 
The former is crucial for triggering 
crystal nucleation \cite{russo_hs}, whereas the latter leads to the frustration against crystallization, a role that increases with polydispersity. None of these physical aspects of the system could be described by two-body quantities. 

We previously showed that spatial fluctuations of crystal-like bond orientational order are closely correlated with local dynamics: 
more ordered regions have slower dynamics \cite{KAT,WT,ShintaniNP,tanaka,Kawasaki3D,mathieu_icosahedra}. 
Together with these results, we may say that it is many-body correlations, or crystal-like bond orientational ordering, that 
are the cause of slow dynamics and dynamic heterogeneity.  
This means that future theories of glass transition and crystallization should deal with many-body correlation effects properly. 
The link between the symmetry of the relevant bond order parameter for describing structural ordering 
in a supercooled liquid and that for crystallization is another interesting point. 
This may be a direct consequence of the fact that glass transition is governed by the same free energy as that 
controlling crystallization \cite{TanakaGJPCM,TanakaJSP,TanakaR}. This conjecture is further supported by the role of 
polydispersity in the glass-forming ability of hard spheres.  

The mechanism by which polydispersity increases the barrier for crystal nucleation may be two fold: 
(i) direct random disorder effect which destroys crystal-like bond orientational order in a supercooled liquid,
which is the first step in crystal nucleation~\cite{russo_hs};
(ii) enhancement of icosahedral ordering with an increase in the polydispersity.
It is known that size disparity between a particle and its surrounding neighbors stabilizes icosahedral order \cite{Shimono2012}. 
Since the symmetry of icosahedral order is not consistent with that of the equilibrium crystal polymorphs (fcc and hcp for this case), 
competing ordering toward these two mutually inconsistent symmetries leads to strong frustration effects on crystallization, as in the case of 
2D spin liquids~\cite{ShintaniNP,STNM}. 
The results shown in Fig. \ref{fig:stability_map} suggest that mechanism (ii) may be more relevant for the suppression of crystallization. 

Although we studied polydisperse hard spheres in this article, these frustration mechanisms should be relevant to many other 
glass-forming systems including metallic glass formers~\cite{Jakse2008,Hwang2012}. 

\section{Summary}\label{sec:summary}

In this article, we show firm evidence for the importance of many-body correlations in glass transition phenomena 
for hard spheres liquids. This feature cannot be described by the standard liquid-state theories based on two-body correlation. 
This implies that, at high density, liquid state packing effects inevitably lead to many-body correlations, which play 
key roles in phenomena like the glass transition and crystallization. 
A physically natural order parameter to pick up these many-body correlations is the \emph{bond order parameter}, whose importance 
has been well recognized and established for ordering transitions of hard disks in 2D \cite{NelsonB}. 
We believe in the importance of incorporating many-body correlations into theories to describe
both the glass transition and crystallization phenomena properly~\cite{TanakaJSP,TanakaR}.

Our study also indicates that there is an intrinsic link between crystallization and vitrification. 
Whether a polydisperse hard spheres system is crystallized or vitrified can be controlled just by changing 
polydispersity, which affects extendable crystal-like bond orientational order and isolated icosahedral order 
in an opposite manner.
We speculate that this direct link may exist for systems where crystallization does not involve phase separation, 
in other words, as far as the two phenomena are described by the same free energy \cite{TanakaGJPCM,TanakaJSP,TanakaR}.  
How universal is this scenario needs to be checked carefully in the future. 

\section*{Acknowledgments}
The authors are grateful to Daniel Bonn for fruitful discussion and drawing our attention to Ref. \cite{Aste2004}. 
This study was partly supported by a grant-in-aid from 
the Ministry of Education, Culture, Sports, Science and Technology, Japan (Kakenhi)
and by the Japan Society for the Promotion of
Science (JSPS) through its ``Funding Program for World-Leading
Innovative R\&D on Science and Technology (FIRST Program)'' and a JSPS Postdoctoral Fellowship.



\begin{thebibliography}{50}
\expandafter\ifx\csname natexlab\endcsname\relax\def\natexlab#1{#1}\fi
\expandafter\ifx\csname bibnamefont\endcsname\relax
  \def\bibnamefont#1{#1}\fi
\expandafter\ifx\csname bibfnamefont\endcsname\relax
  \def\bibfnamefont#1{#1}\fi
\expandafter\ifx\csname citenamefont\endcsname\relax
  \def\citenamefont#1{#1}\fi
\expandafter\ifx\csname url\endcsname\relax
  \def\url#1{\texttt{#1}}\fi
\expandafter\ifx\csname urlprefix\endcsname\relax\def\urlprefix{URL }\fi
\providecommand{\bibinfo}[2]{#2}
\providecommand{\eprint}[2][]{\url{#2}}

\bibitem[{\citenamefont{Petit and Coquerel}(2006)}]{Petit2006}
\bibinfo{author}{\bibfnamefont{S.}~\bibnamefont{Petit}} \bibnamefont{and}
  \bibinfo{author}{\bibfnamefont{G.}~\bibnamefont{Coquerel}}, in
  \emph{\bibinfo{booktitle}{Polymorphism: in the Pharmaceutical Industry}}
  (\bibinfo{publisher}{Wiley-VCH Verlag GmbH \& Co. KGaA},
  \bibinfo{year}{2006}), pp. \bibinfo{pages}{259----285}, ISBN
  \bibinfo{isbn}{9783527607884}.

\bibitem[{\citenamefont{Grzybowska et~al.}(2012)\citenamefont{Grzybowska,
  Paluch, Wlodarczyk, Grzybowski, Kaminski, Hawelek, Zakowiecki, Kasprzycka,
  and Jankowska-Sumara}}]{Grzybowska2012}
\bibinfo{author}{\bibfnamefont{K.}~\bibnamefont{Grzybowska}},
  \bibinfo{author}{\bibfnamefont{M.}~\bibnamefont{Paluch}},
  \bibinfo{author}{\bibfnamefont{P.}~\bibnamefont{Wlodarczyk}},
  \bibinfo{author}{\bibfnamefont{a.}~\bibnamefont{Grzybowski}},
  \bibinfo{author}{\bibfnamefont{K.}~\bibnamefont{Kaminski}},
  \bibinfo{author}{\bibfnamefont{L.}~\bibnamefont{Hawelek}},
  \bibinfo{author}{\bibfnamefont{D.}~\bibnamefont{Zakowiecki}},
  \bibinfo{author}{\bibfnamefont{a.}~\bibnamefont{Kasprzycka}},
  \bibnamefont{and}
  \bibinfo{author}{\bibfnamefont{I.}~\bibnamefont{Jankowska-Sumara}},
  \bibinfo{journal}{Mol. pharm.} \textbf{\bibinfo{volume}{9}},
  \bibinfo{pages}{894} (\bibinfo{year}{2012}), ISSN \bibinfo{issn}{1543-8392},
  \urlprefix\url{http://www.ncbi.nlm.nih.gov/pubmed/22384922}.

\bibitem[{\citenamefont{Cavagna}(2009)}]{cavagna2009supercooled}
\bibinfo{author}{\bibfnamefont{A.}~\bibnamefont{Cavagna}},
  \bibinfo{journal}{Phys. Rep.} \textbf{\bibinfo{volume}{476}},
  \bibinfo{pages}{51} (\bibinfo{year}{2009}), ISSN \bibinfo{issn}{03701573},
  \urlprefix\url{http://linkinghub.elsevier.com/retrieve/pii/S0370157309001112%
}.

\bibitem[{\citenamefont{Berthier and Biroli}(2011)}]{BerthierR}
\bibinfo{author}{\bibfnamefont{L.}~\bibnamefont{Berthier}} \bibnamefont{and}
  \bibinfo{author}{\bibfnamefont{G.}~\bibnamefont{Biroli}},
  \bibinfo{journal}{Rev. Mod. Phys.} \textbf{\bibinfo{volume}{83}},
  \bibinfo{pages}{587} (\bibinfo{year}{2011}).

\bibitem[{\citenamefont{Yamamoto and Onuki}(1998)}]{yamamoto1998}
\bibinfo{author}{\bibfnamefont{R.}~\bibnamefont{Yamamoto}} \bibnamefont{and}
  \bibinfo{author}{\bibfnamefont{A.}~\bibnamefont{Onuki}},
  \bibinfo{journal}{Phys. Rev. E} \textbf{\bibinfo{volume}{58}},
  \bibinfo{pages}{3515} (\bibinfo{year}{1998}).

\bibitem[{\citenamefont{Donati et~al.}(1999)\citenamefont{Donati, Glotzer,
  Poole, Kob, and Plimpton}}]{Donati1999a}
\bibinfo{author}{\bibfnamefont{C.}~\bibnamefont{Donati}},
  \bibinfo{author}{\bibfnamefont{S.~C.} \bibnamefont{Glotzer}},
  \bibinfo{author}{\bibfnamefont{P.~H.} \bibnamefont{Poole}},
  \bibinfo{author}{\bibfnamefont{W.}~\bibnamefont{Kob}}, \bibnamefont{and}
  \bibinfo{author}{\bibfnamefont{S.~J.} \bibnamefont{Plimpton}},
  \bibinfo{journal}{Phys. Rev. E} \textbf{\bibinfo{volume}{60}},
  \bibinfo{pages}{3107} (\bibinfo{year}{1999}), ISSN \bibinfo{issn}{1063-651X},
  \urlprefix\url{http://www.ncbi.nlm.nih.gov/pubmed/11970118}.

\bibitem[{\citenamefont{Tanaka et~al.}(2010)\citenamefont{Tanaka, Kawasaki,
  Shintani, and Watanabe}}]{tanaka}
\bibinfo{author}{\bibfnamefont{H.}~\bibnamefont{Tanaka}},
  \bibinfo{author}{\bibfnamefont{T.}~\bibnamefont{Kawasaki}},
  \bibinfo{author}{\bibfnamefont{H.}~\bibnamefont{Shintani}}, \bibnamefont{and}
  \bibinfo{author}{\bibfnamefont{K.}~\bibnamefont{Watanabe}},
  \bibinfo{journal}{Nature Mater.} \textbf{\bibinfo{volume}{9}},
  \bibinfo{pages}{324} (\bibinfo{year}{2010}).

\bibitem[{\citenamefont{Treacy and Borisenko}(2012)}]{Treacy2012}
\bibinfo{author}{\bibfnamefont{M.~M.~J.} \bibnamefont{Treacy}}
  \bibnamefont{and} \bibinfo{author}{\bibfnamefont{K.~B.}
  \bibnamefont{Borisenko}}, \bibinfo{journal}{Science}
  \textbf{\bibinfo{volume}{335}}, \bibinfo{pages}{950} (\bibinfo{year}{2012}),
  ISSN \bibinfo{issn}{0036-8075},
  \urlprefix\url{http://www.sciencemag.org/cgi/doi/10.1126/science.1214780}.

\bibitem[{\citenamefont{Hwang et~al.}(2012)\citenamefont{Hwang, Melgarejo,
  Kalay, Kalay, Kramer, Stone, and Voyles}}]{Hwang2012}
\bibinfo{author}{\bibfnamefont{J.}~\bibnamefont{Hwang}},
  \bibinfo{author}{\bibfnamefont{Z.}~\bibnamefont{Melgarejo}},
  \bibinfo{author}{\bibfnamefont{Y.}~\bibnamefont{Kalay}},
  \bibinfo{author}{\bibfnamefont{I.}~\bibnamefont{Kalay}},
  \bibinfo{author}{\bibfnamefont{M.}~\bibnamefont{Kramer}},
  \bibinfo{author}{\bibfnamefont{D.}~\bibnamefont{Stone}}, \bibnamefont{and}
  \bibinfo{author}{\bibfnamefont{P.}~\bibnamefont{Voyles}},
  \bibinfo{journal}{Phys. Rev. Lett.} \textbf{\bibinfo{volume}{108}},
  \bibinfo{pages}{195505} (\bibinfo{year}{2012}), ISSN
  \bibinfo{issn}{0031-9007},
  \urlprefix\url{http://link.aps.org/doi/10.1103/PhysRevLett.108.195505}.

\bibitem[{\citenamefont{Lubchenko and Wolynes}(2007)}]{lubchenko2007}
\bibinfo{author}{\bibfnamefont{V.}~\bibnamefont{Lubchenko}} \bibnamefont{and}
  \bibinfo{author}{\bibfnamefont{P.~G.} \bibnamefont{Wolynes}},
  \bibinfo{journal}{Annu. Rev. Phys. Chem.} \textbf{\bibinfo{volume}{58}},
  \bibinfo{pages}{235} (\bibinfo{year}{2007}).

\bibitem[{\citenamefont{Biroli et~al.}(2008)\citenamefont{Biroli, Bouchaud,
  Cavagna, Grigera, and Verrocchio}}]{Biroli2008}
\bibinfo{author}{\bibfnamefont{G.}~\bibnamefont{Biroli}},
  \bibinfo{author}{\bibfnamefont{J.~P.} \bibnamefont{Bouchaud}},
  \bibinfo{author}{\bibfnamefont{a.}~\bibnamefont{Cavagna}},
  \bibinfo{author}{\bibfnamefont{T.~S.} \bibnamefont{Grigera}},
  \bibnamefont{and}
  \bibinfo{author}{\bibfnamefont{P.}~\bibnamefont{Verrocchio}},
  \bibinfo{journal}{Nature Phys.} \textbf{\bibinfo{volume}{4}},
  \bibinfo{pages}{771} (\bibinfo{year}{2008}), ISSN \bibinfo{issn}{1745-2473},
  \urlprefix\url{http://www.nature.com/doifinder/10.1038/nphys1050}.

\bibitem[{\citenamefont{Parisi and Zamponi}(2010)}]{Parisi2010}
\bibinfo{author}{\bibfnamefont{G.}~\bibnamefont{Parisi}} \bibnamefont{and}
  \bibinfo{author}{\bibfnamefont{F.}~\bibnamefont{Zamponi}},
  \bibinfo{journal}{Rev. Mod. Phys.} \textbf{\bibinfo{volume}{82}},
  \bibinfo{pages}{789} (\bibinfo{year}{2010}), ISSN \bibinfo{issn}{0034-6861},
  \urlprefix\url{http://link.aps.org/doi/10.1103/RevModPhys.82.789}.

\bibitem[{\citenamefont{Fasolo and Sollich}(2003)}]{Fasolo2003}
\bibinfo{author}{\bibfnamefont{M.}~\bibnamefont{Fasolo}} \bibnamefont{and}
  \bibinfo{author}{\bibfnamefont{P.}~\bibnamefont{Sollich}},
  \bibinfo{journal}{Phys. Rev. Lett.} \textbf{\bibinfo{volume}{91}},
  \bibinfo{pages}{068301} (\bibinfo{year}{2003}), ISSN
  \bibinfo{issn}{0031-9007},
  \urlprefix\url{http://link.aps.org/doi/10.1103/PhysRevLett.91.068301}.

\bibitem[{\citenamefont{Hopkins et~al.}(2011)\citenamefont{Hopkins, Jiao,
  Stillinger, and Torquato}}]{Hopkins2011b}
\bibinfo{author}{\bibfnamefont{A.}~\bibnamefont{Hopkins}},
  \bibinfo{author}{\bibfnamefont{Y.}~\bibnamefont{Jiao}},
  \bibinfo{author}{\bibfnamefont{F.}~\bibnamefont{Stillinger}},
  \bibnamefont{and} \bibinfo{author}{\bibfnamefont{S.}~\bibnamefont{Torquato}},
  \bibinfo{journal}{Phys. Rev. Lett.} \textbf{\bibinfo{volume}{107}},
  \bibinfo{pages}{125501} (\bibinfo{year}{2011}), ISSN
  \bibinfo{issn}{0031-9007}, \eprint{1108.2210},
  \urlprefix\url{http://arxiv.org/abs/1108.2210
  http://link.aps.org/doi/10.1103/PhysRevLett.107.125501}.

\bibitem[{\citenamefont{Hopkins et~al.}(2012)\citenamefont{Hopkins, Stillinger,
  and Torquato}}]{Hopkins2012}
\bibinfo{author}{\bibfnamefont{A.}~\bibnamefont{Hopkins}},
  \bibinfo{author}{\bibfnamefont{F.}~\bibnamefont{Stillinger}},
  \bibnamefont{and} \bibinfo{author}{\bibfnamefont{S.}~\bibnamefont{Torquato}},
  \bibinfo{journal}{Phys. Rev. E} \textbf{\bibinfo{volume}{85}},
  \bibinfo{pages}{021130} (\bibinfo{year}{2012}), ISSN
  \bibinfo{issn}{1539-3755},
  \urlprefix\url{http://link.aps.org/doi/10.1103/PhysRevE.85.021130}.

\bibitem[{\citenamefont{Charbonneau et~al.}(2012)\citenamefont{Charbonneau,
  Charbonneau, and Tarjus}}]{Charbonneau}
\bibinfo{author}{\bibfnamefont{B.}~\bibnamefont{Charbonneau}},
  \bibinfo{author}{\bibfnamefont{P.}~\bibnamefont{Charbonneau}},
  \bibnamefont{and} \bibinfo{author}{\bibfnamefont{G.}~\bibnamefont{Tarjus}},
  \bibinfo{journal}{Phys. Rev. Lett.} \textbf{\bibinfo{volume}{108}},
  \bibinfo{pages}{035701} (\bibinfo{year}{2012}),
  \urlprefix\url{http://link.aps.org/doi/10.1103/PhysRevLett.108.035701}.

\bibitem[{\citenamefont{Coslovich}(2011)}]{Coslovich2011}
\bibinfo{author}{\bibfnamefont{D.}~\bibnamefont{Coslovich}},
  \bibinfo{journal}{Phys. Rev. E} \textbf{\bibinfo{volume}{83}},
  \bibinfo{pages}{051505} (\bibinfo{year}{2011}), ISSN
  \bibinfo{issn}{1539-3755},
  \urlprefix\url{http://link.aps.org/doi/10.1103/PhysRevE.83.051505}.

\bibitem[{\citenamefont{Malins et~al.}(2012)\citenamefont{Malins, Eggers,
  Royall, Williams, and Tanaka}}]{Malins2012}
\bibinfo{author}{\bibfnamefont{A.}~\bibnamefont{Malins}},
  \bibinfo{author}{\bibfnamefont{J.}~\bibnamefont{Eggers}},
  \bibinfo{author}{\bibfnamefont{C.~P.} \bibnamefont{Royall}},
  \bibinfo{author}{\bibfnamefont{S.~R.} \bibnamefont{Williams}},
  \bibnamefont{and} \bibinfo{author}{\bibfnamefont{H.}~\bibnamefont{Tanaka}}
  (\bibinfo{year}{2012}), \eprint{1203.1732},
  \urlprefix\url{http://arxiv.org/abs/1203.1732}.

\bibitem[{\citenamefont{Mosayebi et~al.}(2010)\citenamefont{Mosayebi, Del~Gado,
  Ilg, and {\"O}ttinger}}]{mosayebi2010}
\bibinfo{author}{\bibfnamefont{M.}~\bibnamefont{Mosayebi}},
  \bibinfo{author}{\bibfnamefont{E.}~\bibnamefont{Del~Gado}},
  \bibinfo{author}{\bibfnamefont{P.}~\bibnamefont{Ilg}}, \bibnamefont{and}
  \bibinfo{author}{\bibfnamefont{H.~C.} \bibnamefont{{\"O}ttinger}},
  \bibinfo{journal}{Phys. Rev. Lett.} \textbf{\bibinfo{volume}{104}},
  \bibinfo{pages}{205704} (\bibinfo{year}{2010}).

\bibitem[{\citenamefont{Mosayebi et~al.}(2012)\citenamefont{Mosayebi, Del~Gado,
  Ilg, and {\"O}ttinger}}]{mosayebi2012}
\bibinfo{author}{\bibfnamefont{M.}~\bibnamefont{Mosayebi}},
  \bibinfo{author}{\bibfnamefont{E.}~\bibnamefont{Del~Gado}},
  \bibinfo{author}{\bibfnamefont{P.}~\bibnamefont{Ilg}}, \bibnamefont{and}
  \bibinfo{author}{\bibfnamefont{H.~C.} \bibnamefont{{\"O}ttinger}},
  \bibinfo{journal}{J. Chem. Phys.} \textbf{\bibinfo{volume}{137}},
  \bibinfo{pages}{024504} (\bibinfo{year}{2012}).

\bibitem[{\citenamefont{Nettleton and Green}(1958)}]{Nettleton1958}
\bibinfo{author}{\bibfnamefont{R.~E.} \bibnamefont{Nettleton}}
  \bibnamefont{and} \bibinfo{author}{\bibfnamefont{M.~S.} \bibnamefont{Green}},
  \bibinfo{journal}{J. Chem. Phys.} \textbf{\bibinfo{volume}{29}},
  \bibinfo{pages}{1365} (\bibinfo{year}{1958}), ISSN \bibinfo{issn}{00219606},
  \urlprefix\url{http://link.aip.org/link/JCPSA6/v29/i6/p1365/s1\&Agg=doi}.

\bibitem[{\citenamefont{Mountain}(1971)}]{Mountain1971}
\bibinfo{author}{\bibfnamefont{R.~D.} \bibnamefont{Mountain}},
  \bibinfo{journal}{J. Chem. Phys.} \textbf{\bibinfo{volume}{55}},
  \bibinfo{pages}{2250} (\bibinfo{year}{1971}), ISSN \bibinfo{issn}{00219606},
  \urlprefix\url{http://link.aip.org/link/?JCP/55/2250/1\&Agg=doi}.

\bibitem[{\citenamefont{Watanabe et~al.}(2011)\citenamefont{Watanabe, Kawasaki,
  and Tanaka}}]{watanabe_walls}
\bibinfo{author}{\bibfnamefont{K.}~\bibnamefont{Watanabe}},
  \bibinfo{author}{\bibfnamefont{T.}~\bibnamefont{Kawasaki}}, \bibnamefont{and}
  \bibinfo{author}{\bibfnamefont{H.}~\bibnamefont{Tanaka}},
  \bibinfo{journal}{Nat. Mater.} \textbf{\bibinfo{volume}{10}},
  \bibinfo{pages}{512} (\bibinfo{year}{2011}).

\bibitem[{\citenamefont{Kawasaki and Tanaka}(2011)}]{KawasakiJPCM}
\bibinfo{author}{\bibfnamefont{T.}~\bibnamefont{Kawasaki}} \bibnamefont{and}
  \bibinfo{author}{\bibfnamefont{H.}~\bibnamefont{Tanaka}},
  \bibinfo{journal}{J. Phys.: Condens. Matter} \textbf{\bibinfo{volume}{23}},
  \bibinfo{pages}{194121} (\bibinfo{year}{2011}).

\bibitem[{\citenamefont{Aste and Coniglio}(2004)}]{Aste2004}
\bibinfo{author}{\bibfnamefont{T.}~\bibnamefont{Aste}} \bibnamefont{and}
  \bibinfo{author}{\bibfnamefont{A.}~\bibnamefont{Coniglio}},
  \bibinfo{journal}{EPL} \textbf{\bibinfo{volume}{67}}, \bibinfo{pages}{165}
  (\bibinfo{year}{2004}), ISSN \bibinfo{issn}{0295-5075},
  \urlprefix\url{http://stacks.iop.org/0295-5075/67/i=2/a=165?key=crossref.294%
04857a396102cb9c207367641c3da}.

\bibitem[{\citenamefont{Steinhardt et~al.}(1983)\citenamefont{Steinhardt,
  Nelson, and Ronchetti}}]{steinhardt}
\bibinfo{author}{\bibfnamefont{P.~J.} \bibnamefont{Steinhardt}},
  \bibinfo{author}{\bibfnamefont{D.~R.} \bibnamefont{Nelson}},
  \bibnamefont{and}
  \bibinfo{author}{\bibfnamefont{M.}~\bibnamefont{Ronchetti}},
  \bibinfo{journal}{Phys. Rev. B} \textbf{\bibinfo{volume}{28}},
  \bibinfo{pages}{784} (\bibinfo{year}{1983}).

\bibitem[{\citenamefont{Auer and Frenkel}(2004)}]{auer}
\bibinfo{author}{\bibfnamefont{S.}~\bibnamefont{Auer}} \bibnamefont{and}
  \bibinfo{author}{\bibfnamefont{D.}~\bibnamefont{Frenkel}},
  \bibinfo{journal}{J. Chem. Phys.} \textbf{\bibinfo{volume}{120}},
  \bibinfo{pages}{3015} (\bibinfo{year}{2004}).

\bibitem[{\citenamefont{Russo and Tanaka}(2012{\natexlab{a}})}]{russo_hs}
\bibinfo{author}{\bibfnamefont{J.}~\bibnamefont{Russo}} \bibnamefont{and}
  \bibinfo{author}{\bibfnamefont{H.}~\bibnamefont{Tanaka}},
  \bibinfo{journal}{Sci. Rep.} \textbf{\bibinfo{volume}{2}},
  \bibinfo{pages}{505} (\bibinfo{year}{2012}{\natexlab{a}}).

\bibitem[{\citenamefont{Lechner and Dellago}(2008)}]{lechner}
\bibinfo{author}{\bibfnamefont{W.}~\bibnamefont{Lechner}} \bibnamefont{and}
  \bibinfo{author}{\bibfnamefont{C.}~\bibnamefont{Dellago}},
  \bibinfo{journal}{J. Chem. Phys.} \textbf{\bibinfo{volume}{129}},
  \bibinfo{eid}{114707} (\bibinfo{year}{2008}).

\bibitem[{\citenamefont{Leocmach and Tanaka}(2012)}]{mathieu_icosahedra}
\bibinfo{author}{\bibfnamefont{M.}~\bibnamefont{Leocmach}} \bibnamefont{and}
  \bibinfo{author}{\bibfnamefont{H.}~\bibnamefont{Tanaka}},
  \bibinfo{journal}{Nat. Commun.} \textbf{\bibinfo{volume}{3}},
  \bibinfo{pages}{974} (\bibinfo{year}{2012}), ISSN \bibinfo{issn}{2041-1723},
  \urlprefix\url{http://www.nature.com/doifinder/10.1038/ncomms1974}.

\bibitem[{\citenamefont{Kapfer et~al.}(2012)\citenamefont{Kapfer, Mickel,
  Mecke, and Schr{\"o}der-Turk}}]{kapfer2012jammed}
\bibinfo{author}{\bibfnamefont{S.}~\bibnamefont{Kapfer}},
  \bibinfo{author}{\bibfnamefont{W.}~\bibnamefont{Mickel}},
  \bibinfo{author}{\bibfnamefont{K.}~\bibnamefont{Mecke}}, \bibnamefont{and}
  \bibinfo{author}{\bibfnamefont{G.}~\bibnamefont{Schr{\"o}der-Turk}},
  \bibinfo{journal}{Phys. Rev. E} \textbf{\bibinfo{volume}{85}},
  \bibinfo{pages}{030301} (\bibinfo{year}{2012}).

\bibitem[{\citenamefont{Kawasaki and Tanaka}(2010{\natexlab{a}})}]{kawasaki}
\bibinfo{author}{\bibfnamefont{T.}~\bibnamefont{Kawasaki}} \bibnamefont{and}
  \bibinfo{author}{\bibfnamefont{H.}~\bibnamefont{Tanaka}},
  \bibinfo{journal}{Proc. Nat. Acad. Sci. U.S.A.}
  \textbf{\bibinfo{volume}{107}}, \bibinfo{pages}{14036}
  (\bibinfo{year}{2010}{\natexlab{a}}), ISSN \bibinfo{issn}{0027-8424}.

\bibitem[{\citenamefont{Russo and Tanaka}(2012{\natexlab{b}})}]{russo_gcm}
\bibinfo{author}{\bibfnamefont{J.}~\bibnamefont{Russo}} \bibnamefont{and}
  \bibinfo{author}{\bibfnamefont{H.}~\bibnamefont{Tanaka}},
  \bibinfo{journal}{Soft Matter} \textbf{\bibinfo{volume}{8}},
  \bibinfo{pages}{4206} (\bibinfo{year}{2012}{\natexlab{b}}).

\bibitem[{\citenamefont{Flenner et~al.}(2011)\citenamefont{Flenner, Zhang, and
  Szamel}}]{Flenner2011}
\bibinfo{author}{\bibfnamefont{E.}~\bibnamefont{Flenner}},
  \bibinfo{author}{\bibfnamefont{M.}~\bibnamefont{Zhang}}, \bibnamefont{and}
  \bibinfo{author}{\bibfnamefont{G.}~\bibnamefont{Szamel}},
  \bibinfo{journal}{Phys. Rev. E} \textbf{\bibinfo{volume}{83}},
  \bibinfo{pages}{051501} (\bibinfo{year}{2011}), ISSN
  \bibinfo{issn}{1539-3755},
  \urlprefix\url{http://link.aps.org/doi/10.1103/PhysRevE.83.051501}.

\bibitem[{not()}]{nota_q6}
\bibinfo{note}{Please note that the calculation in Fourier space is done here
  by retaining the $10\%$ most ordered particles for each order parameter. For
  $q_6$ this means collecting the signal almost entirely from icosahedral
  particles. A real space analysis, done by including all particles, shows
  instead that the correlation length of $q_6$ is increasing as it includes
  also bond-orientational ordered structures.}

\bibitem[{\citenamefont{Shimono and Onodera}(2012)}]{Shimono2012}
\bibinfo{author}{\bibfnamefont{M.}~\bibnamefont{Shimono}} \bibnamefont{and}
  \bibinfo{author}{\bibfnamefont{H.}~\bibnamefont{Onodera}},
  \bibinfo{journal}{Revue de M\'etallurgie} \textbf{\bibinfo{volume}{109}},
  \bibinfo{pages}{41} (\bibinfo{year}{2012}).

\bibitem[{\citenamefont{Zaccarelli et~al.}(2009)\citenamefont{Zaccarelli,
  Valeriani, Sanz, Poon, Cates, and Pusey}}]{zaccarelli}
\bibinfo{author}{\bibfnamefont{E.}~\bibnamefont{Zaccarelli}},
  \bibinfo{author}{\bibfnamefont{C.}~\bibnamefont{Valeriani}},
  \bibinfo{author}{\bibfnamefont{E.}~\bibnamefont{Sanz}},
  \bibinfo{author}{\bibfnamefont{W.~C.~K.} \bibnamefont{Poon}},
  \bibinfo{author}{\bibfnamefont{M.~E.} \bibnamefont{Cates}}, \bibnamefont{and}
  \bibinfo{author}{\bibfnamefont{P.~N.} \bibnamefont{Pusey}},
  \bibinfo{journal}{Phys. Rev. Lett.} \textbf{\bibinfo{volume}{103}},
  \bibinfo{pages}{135704} (\bibinfo{year}{2009}).

\bibitem[{\citenamefont{Pusey et~al.}(2009)\citenamefont{Pusey, Zaccarelli,
  Valeriani, Sanz, Poon, and Cates}}]{pusey2009hard}
\bibinfo{author}{\bibfnamefont{P.}~\bibnamefont{Pusey}},
  \bibinfo{author}{\bibfnamefont{E.}~\bibnamefont{Zaccarelli}},
  \bibinfo{author}{\bibfnamefont{C.}~\bibnamefont{Valeriani}},
  \bibinfo{author}{\bibfnamefont{E.}~\bibnamefont{Sanz}},
  \bibinfo{author}{\bibfnamefont{W.}~\bibnamefont{Poon}}, \bibnamefont{and}
  \bibinfo{author}{\bibfnamefont{M.}~\bibnamefont{Cates}},
  \bibinfo{journal}{Philos. T. R. Soc. A} \textbf{\bibinfo{volume}{367}},
  \bibinfo{pages}{4993} (\bibinfo{year}{2009}).

\bibitem[{\citenamefont{Tanaka}(2003)}]{TanakaK}
\bibinfo{author}{\bibfnamefont{H.}~\bibnamefont{Tanaka}},
  \bibinfo{journal}{Phys. Rev. E} \textbf{\bibinfo{volume}{68}},
  \bibinfo{pages}{011505} (\bibinfo{year}{2003}).

\bibitem[{\citenamefont{Fasolo and Sollich}(2004)}]{fasolo2004fractionation}
\bibinfo{author}{\bibfnamefont{M.}~\bibnamefont{Fasolo}} \bibnamefont{and}
  \bibinfo{author}{\bibfnamefont{P.}~\bibnamefont{Sollich}},
  \bibinfo{journal}{Phys. Rev. E} \textbf{\bibinfo{volume}{70}},
  \bibinfo{pages}{041410} (\bibinfo{year}{2004}).

\bibitem[{\citenamefont{Kawasaki et~al.}(2007)\citenamefont{Kawasaki, Araki,
  and Tanaka}}]{KAT}
\bibinfo{author}{\bibfnamefont{T.}~\bibnamefont{Kawasaki}},
  \bibinfo{author}{\bibfnamefont{T.}~\bibnamefont{Araki}}, \bibnamefont{and}
  \bibinfo{author}{\bibfnamefont{H.}~\bibnamefont{Tanaka}},
  \bibinfo{journal}{Phys. Rev. Lett.} \textbf{\bibinfo{volume}{99}},
  \bibinfo{pages}{215701} (\bibinfo{year}{2007}).

\bibitem[{\citenamefont{Watanabe and Tanaka}(2008)}]{WT}
\bibinfo{author}{\bibfnamefont{K.}~\bibnamefont{Watanabe}} \bibnamefont{and}
  \bibinfo{author}{\bibfnamefont{H.}~\bibnamefont{Tanaka}},
  \bibinfo{journal}{Phys. Rev. Lett.} \textbf{\bibinfo{volume}{100}},
  \bibinfo{pages}{158002} (\bibinfo{year}{2008}).

\bibitem[{\citenamefont{Shintani and Tanaka}(2006)}]{ShintaniNP}
\bibinfo{author}{\bibfnamefont{H.}~\bibnamefont{Shintani}} \bibnamefont{and}
  \bibinfo{author}{\bibfnamefont{H.}~\bibnamefont{Tanaka}},
  \bibinfo{journal}{Nature Phys.} \textbf{\bibinfo{volume}{2}},
  \bibinfo{pages}{200} (\bibinfo{year}{2006}).

\bibitem[{\citenamefont{Kawasaki and Tanaka}(2010{\natexlab{b}})}]{Kawasaki3D}
\bibinfo{author}{\bibfnamefont{T.}~\bibnamefont{Kawasaki}} \bibnamefont{and}
  \bibinfo{author}{\bibfnamefont{H.}~\bibnamefont{Tanaka}},
  \bibinfo{journal}{J. Phys.: Condens. Matter} \textbf{\bibinfo{volume}{22}},
  \bibinfo{pages}{232102} (\bibinfo{year}{2010}{\natexlab{b}}).

\bibitem[{\citenamefont{Tanaka}(2010)}]{TanakaJSP}
\bibinfo{author}{\bibfnamefont{H.}~\bibnamefont{Tanaka}}, \bibinfo{journal}{J.
  Stat. Mech.} \textbf{\bibinfo{volume}{2010}}, \bibinfo{pages}{P12001}
  (\bibinfo{year}{2010}).

\bibitem[{\citenamefont{Tanaka}(2012)}]{TanakaR}
\bibinfo{author}{\bibfnamefont{H.}~\bibnamefont{Tanaka}},
  \bibinfo{journal}{Eur. Phys. J. E} \textbf{\bibinfo{volume}{35}},
  \bibinfo{pages}{113} (\bibinfo{year}{2012}).

\bibitem[{\citenamefont{Tanaka}(1998)}]{TanakaGJPCM}
\bibinfo{author}{\bibfnamefont{H.}~\bibnamefont{Tanaka}}, \bibinfo{journal}{J.
  Phys.: Condens. Matter} \textbf{\bibinfo{volume}{10}}, \bibinfo{pages}{L207}
  (\bibinfo{year}{1998}).

\bibitem[{\citenamefont{Shintani and Tanaka}(2008)}]{STNM}
\bibinfo{author}{\bibfnamefont{H.}~\bibnamefont{Shintani}} \bibnamefont{and}
  \bibinfo{author}{\bibfnamefont{H.}~\bibnamefont{Tanaka}},
  \bibinfo{journal}{Nature Mater.} \textbf{\bibinfo{volume}{7}},
  \bibinfo{pages}{870} (\bibinfo{year}{2008}).

\bibitem[{\citenamefont{Jakse and Pasturel}(2008)}]{Jakse2008}
\bibinfo{author}{\bibfnamefont{N.}~\bibnamefont{Jakse}} \bibnamefont{and}
  \bibinfo{author}{\bibfnamefont{A.}~\bibnamefont{Pasturel}},
  \bibinfo{journal}{Phys. Rev. B} \textbf{\bibinfo{volume}{78}},
  \bibinfo{pages}{214204} (\bibinfo{year}{2008}).

\bibitem[{\citenamefont{Nelson}(2002)}]{NelsonB}
\bibinfo{author}{\bibfnamefont{D.~R.} \bibnamefont{Nelson}},
  \emph{\bibinfo{title}{Defects and Geometry in Condensed Matter Physics}}
  (\bibinfo{publisher}{Cambridge University Press., Cambridge},
  \bibinfo{year}{2002}).

\end{thebibliography}

\end{document}